\documentclass[12pt,oneside]{article}
\usepackage{amsmath,amssymb,amsfonts,amsthm}
\usepackage{color}

\newcommand{\prof}{\noindent \textit{\textbf{Proof.\:\:}}}

\newcommand\overcirc[1]{\raisebox{10pt}{\tiny$\circ$}{\kern-6.5pt}\mbox{$#1$}}
\newcommand\undersym[2]{\raisebox{-6pt}{\tiny$#2$}{\kern-5pt}\mbox{$#1$}}

\def\Section#1{\vspace{30truept}\addtocounter{section}{1}\setcounter{thm}{0}\setcounter{equation}{0}
{\noindent\Large\bf\arabic{section}.~~#1}\par \vspace{12pt}}

\newtheorem{thm}{Theorem}[section]
\newtheorem{cor}[thm]{Corollary}

\newtheorem{prop}[thm]{Proposition}
\newtheorem{defn}[thm]{Definition}

\newtheorem{rem}[thm]{Remark}

\numberwithin{equation}{section}

\title{ \bf LINEAR CONNECTIONS AND CURVATURE TENSORS IN THE GEOMETRY OF PARALLELIZABLE MANIFOLDS}
\author{\textbf{Nabil L. Youssef \, and \, Amr M.  Sid-Ahmed}}
\date{}

\begin{document}               
\bibliographystyle{plain}
\maketitle                     
\vspace{-1.15cm}
\begin{center}
{Department of Mathematics, Faculty of Science,\\ Cairo University,
Giza, Egypt.}
\end{center}

\vspace{-0.8cm}
\begin{center}
nyoussef@frcu.eun.eg \,\,and\,\, amrs@mailer.eun.eg
\end{center}

\begin{center}
{\textbf{Dedicated to the memory of Prof. Dr. A. TAMIM}}
\end{center}
\smallskip
\bigskip
\noindent{\bf Abstract.} In this paper we discuss linear connections
and curvature tensors in the context of geometry of parallelizable
manifolds (or absolute parallelism geometry). Different curvature
tensors are expressed in a compact form in terms of the torsion
tensor of the canonical connection. Using the Bianchi identities,
some other identities are derived from the expressions obtained.
These identities, in turn, are used to reveal some of the properties
satisfied by an intriguing fourth order tensor which we refer to as
Wanas tensor.\newline A further condition on the canonical
connection is imposed, assuming it is semi-symmetric. The formulae
thus obtained, together with other formulae (Ricci tensors and
scalar curvatures of the different connections admitted by the
space) are calculated under this additional assumption. Considering
a specific form of the semi-symmetric connection causes all
nonvanishing curvature tensors to coincide, up to a constant, with
the Wanas tensor.\newline Physical aspects of some of the geometric
objects considered are pointed out. \footnote {This paper was
presented in \lq\lq\,The international Conference on Developing and
Extending Einstein's Ideas\,\rq\rq held at Cairo, Egypt, December
19-21, 2005.}

\bigskip
\medskip\noindent{\bf Keywords:\/}\, Parallelizable manifold,
Absolute parallelism geometry, dual connection, semi-symmetric
connection, Wanas tensor, Field equations.

\bigskip
\medskip\noindent{\bf  AMS Subject Classification.\/} 51P05, 83C05, 53B21.

\newpage

\Section{Introduction} After the success of his general theory of
gravitation (GR), Einstein searched for a more general theory that
would appropriately describe electromagnetic phenomena together with
gravity. His search for such a unified theory led him to consider
Absolute Parallelism (AP-) geometry \cite{En}. The reason for this
is that AP-geometry is wider than the standard Riemannian geometry.
According to GR, the ten degrees of freedom (the metric components
for $n = 4$) are just sufficient to describe gravitational phenomena
alone. AP-geometry, on the other hand, has sixteen degrees of
freedom. Riemannian geometry is thus relatively limited compared to
AP-geometry which has six extra degrees of freedom. These extra
degrees of freedom could be used to describe physical phenomena
other than gravity. This idea of increasing the number of degrees of
freedom from ten to sixteen is another alternative to the idea of
increasing the dimension of the underlying manifold in the so-called
Kaluza-Klein theory. Moreover, as opposed to Riemannian geometry
which admits only one symmetric affine connection and hence only one
curvature tensor, the AP-space admits at least four built-in
(natural) affine connections, two of which are non-symmetric.
AP-geometry also admits tensors of third order, a number of second
order skew and symmetric tensors and a non-vanishing torsion. These
extra geometric structures, which have no counterpart in Riemannian
geometry, make AP-geometry much richer in its content and hence a
potential candidate for geometric unification schemes. A further
advantage is that associated to an AP-space there is a Riemannian
structure defined in a natural way; thus the AP-space contains
within its geometric structure all the mathematical machinery of
Riemannian geometry. This facilitates comparison between the results
obtained in the context of AP-geometry with the classical GR based
on Riemannian geometry. Finally, calculations within the framework
of AP-geometry are relatively easier than those used in the context
of Riemannian geometry.

\bigskip

In the present paper we investigate the curvature tensors
corresponding to the different natural connections defined in an
AP-space. The paper is organized in the following manner. In section
2, we give a brief account of the basic elements of AP-geometry; we
focus our attention on the fundamental concepts that will be needed
in the sequel. In section 3, we consider the curvature tensors of
the dual and symmetric connections associated to the canonical
connection $\Gamma$ together with the Riemannian curvature. Simple
and compact expressions for such curvature tensors, in terms of the
torsion of $\Gamma$, are deduced. We then use the Bianchi identities
to derive some further interesting identities which simplify the
formulae thus obtained.  In section 4, we study some of the
properties satisfied by an intriguing fourth order tensor, which we
call {\it Wanas tensor} (or simply W-tensor) after M.I. Wanas who
first defined and used such a tensor \cite{a} . The W-tensor is
shown to have  some properties which are similar to those of the
Riemannian curvature tensor. In section 5, we further study the
consequences of assuming that the canonical connection is
semi-symmetric. All curvature tensors with their associated
contractions (Ricci tensors and scalar curvatures) are then derived
in this case. We next consider an interesting special case which
considerably simplifies the formulae thus obtained and show that all
nonvanishing curvature tensors admitted by the space coincide (up to
a constant factor) with the W-tensor. Physical aspects or physical
interpretations of some geometric objects considered are pointed
out.

\Section{A brief account of AP-geometry}

In this section, we give a brief account of the geometry of
parallelizable manifolds (or absolute parallelism geometry). For
more details concerning this geometry, we refer for example to
\cite{HP}, \cite{a}, \cite{b} and \cite{BC}. Some other works
related the subject are \cite{Tam1} and \cite{Tam2}.

A parallelizable manifold \cite{BC} is an n-dimensional $C^{\infty}$
manifold $M$ which admits $n$ linearly independent global vector
fields $\undersym{\lambda}{i}$ $(i = 1, ..., n)$ on $M$. Such a
space is also known in the literature as absolute parallelism space
(AP-space). We will rather use the expressions {\lq\lq {AP-space}
\rq\rq} and {\lq\lq {AP-geometry} \rq\rq} for their typographical
simplicity.

Let \ $\undersym{\lambda}{i}^{\mu}$ $(\mu = 1, ..., n)$ be the
coordinate components of the i-th vector field
$\undersym{\lambda}{i}$. The Einestein summation convention is
applied on both Latin (mesh) and Greek (world) indices, where all
Latin indices are written as subscripts. In the sequel, we will
simply use the symbol $\lambda$ (without a mesh index) to denote any
one of the vector fields $ \, \undersym{\lambda}{i}$ ($i=1 ... ,n$)
and, in most cases,
 when mesh indices appear they will be in pairs, meaning summation.
 The covariant components of \ ${\lambda}^{\mu}$ are
given via the relations\vspace{-0.2cm}
$$\undersym{\lambda}{i}^{\mu}\,\undersym{\lambda}{i}_\nu = \delta^{\mu}_{\nu}, \ \
\undersym{\lambda}{i}^{\mu}\,\undersym{\lambda}{j}_{\mu} =
\delta_{ij}.\vspace{-0.2cm}$$ The $n^{3}$ functions
$\Gamma^{\alpha}_{\mu\nu}$ defined by\vspace{-0.2cm}
$$\Gamma^{\alpha}_{\mu\nu}: =
\undersym{\lambda}{i}^{\alpha}\,\undersym{\lambda}{i}_{\mu,\nu}\vspace{-0.2cm}$$
transform as the coefficients of a linear connection under a change
of coordinates (where the comma denotes partial differentiation with
respect to the coordinate functions). The connection
$\Gamma^{\alpha}_{\mu\nu}$ is clearly non-symmetric and is referred
to as the canonical connection of the space. As easily checked, we
have\vspace{-0.2cm}
$${\lambda}_{\mu|\nu} = 0, \ \ {\lambda}^{\mu}\,_{|\nu} = 0,\vspace{-0.2cm}$$
where the stroke denotes covariant differentiation with respect to
the canonical connection $\Gamma^{\alpha}_{\mu\nu}$. The above
relation is known in the literature as the condition of absolute
parallelism.

Let $\Lambda^{\alpha}_{\mu\nu}: = \Gamma^{\alpha}_{\mu\nu} -
\Gamma^{\alpha}_{\nu\mu}$ denote the torsion tensor of
$\Gamma^{\alpha}_{\mu\nu}$. It is of particular importance to note
that the condition of absolute parallelism together with the
commutation formula
$${\lambda}^{\alpha} \,_{|\mu\nu} - {\lambda}^{\alpha} \,_{|\nu\mu} =
{\lambda}^{\epsilon}R^{\alpha}_{\epsilon\mu\nu} + {\lambda}^{\alpha}
\,_{|\epsilon}\, \Lambda^{\epsilon}_{\nu\mu}$$ forces the curvature
tensor $R^{\alpha}_{\mu\nu\sigma}$ of the canonical connection
$\Gamma^{\alpha}_{\mu\nu}$ to vanish identically. (This is a simple
way, which needs no complicated calculations, to prove the vanishing
of the curvature tensor $R^{\alpha}_{\mu\nu\sigma}$). It is for this
reason that many authors think that the AP-space is a flat space.
This is by no means true. In fact, it is meaningless to speak of
curvature without reference to a connection. All we can say is that
the AP-space is flat with respect to its canonical connection, or
that its canonical connection is flat. However, there are other
three natural (built-in) connections which are nonflat. Namely, the
dual connection
$$\widetilde{{\Gamma}}^{\alpha}_{\mu\nu}: = \Gamma^{\alpha}_{\nu\mu},\vspace{-0.2cm}$$
the symmetric connection\vspace{-0.2cm}
$$\widehat{{\Gamma}}^{\alpha}_{\mu\nu}: =
\frac{1}{2}(\Gamma^{\alpha}_{\mu\nu} + \Gamma^{\alpha}_{\nu\mu}) =
\Gamma^{\alpha}_{(\mu\nu)}\vspace{-0.2cm}$$ and the Riemannian
connection (Christoffel symbols)
$$\overcirc{{\Gamma}}^{\alpha}_{\mu\nu}: = \frac{1}{2}\{g^{\alpha\epsilon}(g_{\epsilon\nu, \mu} +
g_{\epsilon\mu, \nu} - g_{\mu\nu, \epsilon})\}$$ associated to the
metric structure defined by\vspace{-0.2cm}
$$g_{\mu\nu}: = \undersym{\lambda}{i}_{\mu}\,\,\undersym{\lambda}{i}_{\nu}\vspace{-0.2cm}$$
with inverse\vspace{-0.1cm}
$$g^{\mu\nu} = \undersym{\lambda}{i}^{\mu}\,\,\undersym{\lambda}{i}^{\nu}.\vspace{-0.1cm}$$
It is to be noted that the condition of absolute parallelism implies
that the canonical connection $\Gamma^{\alpha}_{\mu\nu}$ is
metric\,:\vspace{-0.1cm}
$$g_{\mu\nu|\sigma} = 0, \ \ g^{\mu\nu} \ _{|\sigma} = 0.\vspace{-0.1cm}$$
Consequently, the covariant differentiation defined by the canonical
connection commutes with contraction by the metric tensor
$g_{\mu\nu}$ and its inverse $g^{\mu\nu}$.

The contortion tensor is defined by\vspace{-0.2cm}
$$\gamma^{\alpha}_{\mu\nu}: = \Gamma^{\alpha}_{\mu\nu} - \overcirc{\Gamma}^{\alpha}_{\mu\nu}.\vspace{-0.2cm}$$
Since \ $\overcirc{\Gamma}^{\alpha}_{\mu\nu}$ is symmetric, it
follows that\vspace{-0.2cm}
$$\Lambda^{\alpha}_{\mu\nu} = \gamma^{\alpha}_{\mu\nu} - \gamma^{\alpha}_{\nu\mu}.\vspace{-0.2cm}$$
Moreover, it can be shown that \vspace{-0.2cm}
$$\gamma^{\alpha}_{\mu\nu} = \undersym{\lambda}{i}^{\alpha}\,\,
\undersym{\lambda}{i}_{\mu{o\atop |}\nu},\vspace{-0.2cm}$$ where
\lq\lq\,$\overcirc\atop|$\rq\rq\, denotes covariant differentiation
with respect to the Riemannian connection \
$\overcirc{\Gamma}^{\alpha}_{\mu\nu}.$ Finally, the contortion
tensor can be expressed in terms of the torsion tensor in the form
\cite{HS}:\vspace{-0.2cm}
$$\gamma_{\mu\nu\sigma} = \frac{1}{2}(\Lambda_{\mu\nu\sigma} +\Lambda_{\sigma\nu\mu} +
\Lambda_{\nu\sigma\mu}),$$ where $\gamma_{\mu\nu\sigma} =
g_{\epsilon\mu}\gamma^{\epsilon}_{\nu\sigma}$ and
$\Lambda_{\mu\nu\sigma} =
g_{\epsilon\mu}\Lambda^{\epsilon}_{\nu\sigma}$. It is to be noted
that $\Lambda_{\mu\nu\sigma}$ is skew-symmetric in the last pair of
indices, whereas $\gamma_{\mu\nu\sigma}$ is skew-symmetric in the
first pair of indices. It is to be noted also that the contortion
tensor vanishes if and only if the torsion tensor vanishes.

We have four types of covariant derivatives corresponding to the
four connections  mentioned above, namely
$$A^{\mu}\,_{|\nu} = A^{\mu}\,_{,\nu} + \Gamma^{\mu}_{\epsilon\nu}\,A^{\epsilon},$$
$$A^{\mu}\,_{\widetilde{{|}}\nu} = A^{\mu}\,_{,\nu} + \widetilde{{\Gamma}}^{\mu}_{\epsilon\nu}\, A^{\epsilon},$$
$$A^{\mu}\,_{\widehat{{|}}\nu} = A^{\mu}\,_{,\nu} + \widehat{{\Gamma}}^{\mu}_{\epsilon\nu}\, A^{\epsilon},$$
$${A^{\mu}}_{{o\atop |}\nu} = A^{\mu}\,_{,\nu} + \overcirc{\Gamma}^{\mu}_{\epsilon\nu}\,A^{\epsilon},$$
where $A^{\mu}$ is an arbitrary contravariant vector.

In conclusion, the AP-space has four curvature tensors
$R^{\alpha}_{\mu\nu\sigma}$,
$\widetilde{{R}}^{\alpha}_{\mu\nu\sigma}$,
$\widehat{{R}}^{\alpha}_{\mu\nu\sigma}$ and \
$\overcirc{R}^{\alpha}_{\mu\nu\sigma}$ corresponding to the four
connections $\Gamma^{\alpha}_{\mu\nu}$,
$\widetilde{{\Gamma}}^{\alpha}_{\mu\nu}$,
$\widehat{{\Gamma}}^{\alpha}_{\mu\nu}$ and \
$\overcirc{\Gamma}^{\alpha}_{\mu\nu}$ respectively. As already
mentioned, only one of these curvature tensors vanishes identically
(the curvature $R^{\alpha}_{\mu\nu\sigma}$ of the canonical
connection ${\Gamma}^{\alpha}_{\mu\nu}$). The other three do not
vanish in general. The vanishing of $R^{\alpha}_{\mu\nu\sigma}$
enables us to express the other three curvature tensors in terms of
the torsion tensor $\Lambda^{\alpha}_{\mu\nu}$ as it will be
revealed in the next section.
\\

\begin{center} {\bf Summary of the geometry of the AP-space}\\[0.3cm]
\begin{tabular}
{|c|c|c|c|c|c|c|c|c|c|c|c|}\hline
&&&&&\\
 Connection&Coefficients&Covariant
 &Torsion&Curvature&Metricity\\
 &&derivative&&&\\[0.2cm]\hline
 &&&&&\\
 Canonical&$\Gamma^{\alpha}_{\mu\nu}$&$|$
 &$\Lambda^{\alpha}_{\mu\, \nu}$&0&metric\\[0.2cm]\hline
 &&&&&\\
Dual&$\widetilde{{\Gamma}}^{\alpha}_{\mu\nu}$&$\begin{array}{cc}\widetilde {}\\[-0.3cm]|\end{array}$
 &$-\Lambda^{\alpha}_{\mu\, \nu}$&$\widetilde{R}^{\alpha}_{\mu\nu\sigma}$&non-metric
 \\[0.2cm]\hline
  &&&&&\\
Symmetric&$\widehat{{\Gamma}}^{\alpha}_{\mu\nu}$&$\begin{array}{cc}
\widehat {}\\[-0.3cm]|\end{array}$&0&$\widehat{R}^{\alpha}_{\mu\nu\sigma}$&non-metric\\[0.2cm]\hline
 &&&&&\\
Riemannian&$\overcirc{\Gamma}^{\alpha}_{\mu\nu}$&$\overcirc\atop
|$&0&$\overcirc{R}^{\alpha}_{\mu\nu\sigma}$&metric\\[0.2cm]\hline
\end{tabular}
\end{center}

\vspace{0.5cm}
\Section{Curvature tensors and the Bianchi identities in\vspace{7pt}
AP-geometry}
 Let $(M,\lambda)$ be an AP-space of
dimension $n$, where $\lambda$ denotes any one of the $n$ linearly
independent vector fields $\,\undersym\lambda{i}\,(i = 1, ..., n)$
defining the AP-structure on $M$. Let $\Gamma^{\alpha}_{\mu\nu}$ be
the canonical connection on $M$ defined by $\Gamma^{\alpha}_{\mu\nu}
= \undersym{\lambda}{i}^{\alpha}\,\undersym{\lambda}{i}_{\mu,\nu}$.
Let $\widetilde{\Gamma}^{\alpha}_{\mu\nu}$, $\widehat
{\Gamma}^{\alpha}_{\mu\nu}$ and \
$\overcirc{\Gamma}^{\alpha}_{\mu\nu}$ be respectively the dual
connection associated to $\Gamma^{\alpha}_{\mu\nu}$, the symmetric
connection associated to $\Gamma^{\alpha}_{\mu\nu}$ and the
Riemannian connection defined by the metric tensor $g_{\mu\nu} =
\undersym{\lambda}{i}_{\mu}\,\,\undersym{\lambda}{i}_{\nu}.$ As in
the previous section, covariant differentiation with respect to
$\Gamma^{\alpha}_{\mu\nu}$, $\widetilde {\Gamma}^{\alpha}_{\mu\nu}$
$\widehat {\Gamma}^{\alpha}_{\mu\nu}$ and \
$\overcirc{\Gamma}^{\alpha}_{\mu\nu}$ will be denoted by $|$,
$\widetilde {|}$, $\widehat {|}$ and $\overcirc \atop |$
respectively.

\begin{thm} The curvature tensors $\widetilde{R}^{\alpha}_{\mu\nu\sigma}$,
$\widehat {R}^{\alpha}_{\mu\nu\sigma}$ and \
$\overcirc{R}^{\alpha}_{\mu\nu\sigma}$ of the connections
$\widetilde{\Gamma}^{\alpha}_{\mu\nu}$,
$\widehat{\Gamma}^{\alpha}_{\mu\nu}$ and \
$\overcirc{\Gamma}^{\alpha}_{\mu\nu}$ are expressed in terms of the
torsion tensor $\Lambda^{\alpha}_{\mu\nu}$ of the canonical
connection $\Gamma^{\alpha}_{\mu\nu}$ as follows:
\begin{description}
\item[(a)] $\widetilde{R}^{\alpha}_{\mu\nu\sigma} = \Lambda^{\alpha}_{\sigma\nu|\mu}.$
\item[(b)] $\widehat {R}^{\alpha}_{\mu\nu\sigma} =
\frac{1}{2}(\Lambda^{\alpha}_{\mu\nu|\sigma} -
\Lambda^{\alpha}_{\mu\sigma|\nu}) +
\frac{1}{4}(\Lambda^{\epsilon}_{\mu\nu}\Lambda^{\alpha}_{\sigma\epsilon}
-
 \Lambda^{\epsilon}_{\mu\sigma}\Lambda^{\alpha}_{\nu\epsilon}) +
\frac{1}{2}(\Lambda^{\epsilon}_{\sigma\nu}\Lambda^{\alpha}_{\epsilon\mu}).$
\item[(c)] $\overcirc{R}^{\alpha}_{\mu\nu\sigma} = (\gamma^{\alpha}_{\mu\nu|\sigma} - \gamma^{\alpha}_{\mu\sigma|\nu}) +
(\gamma^{\epsilon}_{\mu\sigma}\gamma^{\alpha}_{\epsilon\nu} -
\gamma^{\epsilon}_ {\mu\nu}\gamma^{\alpha}_{\epsilon\sigma}) +
\gamma^{\alpha}_{\mu\epsilon}\Lambda^{\epsilon}_{\nu\sigma}.$
\end{description}
\end{thm}

\prof  We start by proving the first relation:
\begin{eqnarray*}\widetilde {R}^{\alpha}_{\mu\nu\sigma} &=&
\widetilde {\Gamma}^{\alpha}_{\mu\sigma,\nu} - \widetilde
{\Gamma}^{\alpha}_{\mu\nu,\sigma} + \widetilde
{\Gamma}^{\epsilon}_{\mu\sigma}\widetilde
{\Gamma}^{\alpha}_{\epsilon\nu} - \widetilde
{\Gamma}^{\epsilon}_{\mu\nu}\widetilde {\Gamma}^{\alpha}_
{\epsilon\sigma}\\
&=& {\Gamma}^{\alpha}_{\sigma\mu,\nu} -
{\Gamma}^{\alpha}_{\nu\mu,\sigma} +
{\Gamma}^{\epsilon}_{\sigma\mu}{\Gamma}^{\alpha}_{\nu\epsilon} -
{\Gamma}^{\epsilon}_{\nu\mu}{\Gamma}^{\alpha}_{\sigma\epsilon}\\
&=&\{\Gamma^{\alpha}_{\sigma\mu,\nu} + \Gamma^{\epsilon}_{\sigma\mu}
(\Lambda^{\alpha}_{\nu\epsilon} + \Gamma^{\alpha}_{\nu\epsilon})\} -
\{\Gamma^{\alpha}_{\nu\mu,\sigma} +
\Gamma^{\epsilon}_{\nu\mu}(\Lambda^{\alpha}_
{\sigma\epsilon} + \Gamma^{\alpha}_{\epsilon\sigma})\}\\
&=& (\Gamma^{\alpha}_{\sigma\mu,\nu} + \Gamma^{\epsilon}_
{\sigma\mu}\Gamma^{\alpha}_{\epsilon\nu}) -
(\Gamma^{\alpha}_{\nu\mu,\sigma} +
\Gamma^{\epsilon}_{\nu\mu}\Gamma^{\alpha}_{\epsilon\sigma}) -
(\Gamma^{\epsilon}_ {\sigma\mu}\Lambda^{\alpha}_{\epsilon\nu} +
\Gamma^{\epsilon}_{\nu\mu}\Lambda
^{\alpha}_{\sigma\epsilon})\\
&=& (R^{\alpha}_{\sigma\nu\mu} + \Gamma^{\alpha}_{\sigma\nu,\mu} +
\Gamma^{\epsilon}_{\sigma\nu}\Gamma^{\alpha}_{\epsilon\mu}) +
(R^{\alpha}_{\nu\mu\sigma} - \Gamma^{\alpha}_{\nu\sigma,\mu} -
\Gamma^{\epsilon}_{\nu\sigma}\Gamma^{\alpha}_{\epsilon\mu}) -
(\Gamma^{\epsilon}_{\sigma\mu}\Lambda^{\alpha}_{\epsilon\nu} +
\Gamma^{\epsilon}_{\nu\mu}\Lambda^{\alpha}_{\sigma\epsilon}).
\end{eqnarray*}
Taking into account the fact that $R^{\alpha}_{\mu\nu\sigma} = 0$,
we get\vspace{-0.2cm}
$$\widetilde{R}^{\alpha}_{\mu\nu\sigma} = \Lambda^{\alpha}_{\sigma\nu,\mu} +
\Gamma^{\alpha}_{\epsilon\mu}\Lambda^{\epsilon}_{\sigma\nu} -
\Gamma^{\epsilon}_{\sigma\mu}\Lambda^{\alpha}_{\epsilon\nu} -
\Gamma^{\epsilon}_{\nu\mu}\Lambda^{\alpha}_{\sigma\epsilon} =
\Lambda^{\alpha}_{\sigma\nu|\mu}.\vspace{-0.2cm}$$

We then prove the second relation: We have, by
definition,\vspace{-0.2cm}
$$\widehat {R}^{\alpha}_{\mu\nu\sigma} = \widehat {\Gamma}^{\alpha}_{\mu\sigma,\nu} -
\widehat {\Gamma}^{\alpha}_{\mu\nu,\sigma} + \widehat
{\Gamma}^{\epsilon}_ {\mu\sigma}\widehat
{\Gamma}^{\alpha}_{\epsilon\nu} - \widehat
{\Gamma}^{\epsilon}_{\mu\nu} \widehat
{\Gamma}^{\alpha}_{\epsilon\sigma}.\vspace{-0.2cm}$$
Now,\vspace{-0.2cm}
\begin{eqnarray*}
\widehat {\Gamma}^{\epsilon}_{\mu\sigma}\widehat
{\Gamma}^{\alpha}_{\epsilon\nu} &=&
\frac{1}{4}(\Lambda^{\epsilon}_{\sigma\mu} +
2\Gamma^{\epsilon}_{\mu\sigma})
(\Lambda^{\alpha}_{\nu\epsilon} + 2\Gamma^{\alpha}_{\epsilon\nu})\\
&=& -
\frac{1}{4}\Lambda^{\epsilon}_{\mu\sigma}\Lambda^{\alpha}_{\nu\epsilon}
-
\frac{1}{2}\Lambda^{\epsilon}_{\mu\sigma}\Gamma^{\alpha}_{\epsilon\nu}
-
\frac{1}{2}\Lambda^{\alpha}_{\epsilon\nu}\Gamma^{\epsilon}_{\mu\sigma}
+
\Gamma^{\epsilon}_{\mu\sigma}\Gamma^{\alpha}_{\epsilon\nu}.\\
\end{eqnarray*}
\\[ - 1.2 cm] Similarly,
$$\widehat {\Gamma}^{\epsilon}_{\mu\nu}\widehat {\Gamma}^{\alpha}_{\epsilon\sigma} =
-\frac{1}{4}\Lambda^{\epsilon}_{\mu\nu}\Lambda^{\alpha}_{\sigma\epsilon}
-
\frac{1}{2}\Lambda^{\epsilon}_{\mu\nu}\Gamma^{\alpha}_{\epsilon\sigma}
-
\frac{1}{2}\Lambda^{\alpha}_{\epsilon\sigma}\Gamma^{\epsilon}_{\mu\nu}
+ \Gamma^{\epsilon}_{\mu\nu}\Gamma^{\alpha}_{\epsilon\sigma}.$$
Moreover,
$$\widehat {\Gamma}^{\alpha}_{\mu\sigma,\nu} =
- \frac{1}{2}\Lambda^{\alpha}_{\mu\sigma,\nu} +
\Gamma^{\alpha}_{\mu\sigma,\nu}\ \text {\, and \, } \ \widehat
{\Gamma}^{\alpha}_{\mu\nu,\sigma} = -
\frac{1}{2}\Lambda^{\alpha}_{\mu\nu,\sigma} +
\Gamma^{\alpha}_{\mu\nu,\sigma}.$$ Hence, noting that
$R^{\alpha}_{\mu\nu\sigma} = 0$, we get\vspace{-0.2cm}
\begin{eqnarray*}\widehat {R}^{\alpha}_{\mu\nu\sigma} &=&
\frac{1}{4}(\Lambda^{\epsilon}_{\mu\nu}\Lambda^{\alpha}_{\sigma\epsilon}
- \Lambda^{\epsilon}_{\mu\sigma}\Lambda^{\alpha}_{\nu\epsilon})
 + \frac{1}{2}\{(\Lambda^{\alpha}_{\mu\nu,\sigma} + \Lambda^{\epsilon}_{\mu\nu}
\Gamma^{\alpha}_{\epsilon\sigma} -
\Lambda^{\alpha}_{\epsilon\nu}\Gamma^{\epsilon}_
{\mu\sigma})\\
&& - \ (\Lambda^{\alpha}_{\mu\sigma,\nu} + \Lambda^{\epsilon}_{\mu\sigma}
\Gamma^{\alpha}_{\epsilon\nu} - \Lambda^{\alpha}_{\epsilon\sigma}
\Gamma^{\epsilon}_{\mu\nu})\}\\
&=& \frac{1}{2}(\Lambda^{\alpha}_{\mu\nu|\sigma} -
\Lambda^{\alpha}_{\mu\sigma|\nu}) +
\frac{1}{4}(\Lambda^{\epsilon}_{\mu\nu}\Lambda^{\alpha}_{\sigma\epsilon}
- \Lambda^{\epsilon}_{\mu\sigma}\Lambda^{\alpha}_{\nu\epsilon})
 + \frac{1}{2}(\Lambda^{\epsilon}_{\sigma\nu}\Lambda^{\alpha}_{\epsilon\mu}).\vspace{-0.2cm}
\end{eqnarray*}
 The proof of relation (c) is carried out in the same manner and
we omit it.\ \ $\Box$
\begin{rem} The first and second formulae {\em({\it resp. The third formula})} of the above theorem
remain {\em({\it resp. remains})} valid in the more general context
in which $\Gamma^{\alpha}_{\mu\nu}$ is any given non-symmetric
linear connection on a manifold $M$ {\em({\it resp. a Riemannian
manifold $(M, g)$})} with vanishing curvature.
\end{rem}
It is clear that the torsion tensor plays the key role in all
identities obtained above. The vanishing of the torsion tensor
forces the three connections $\Gamma^{\alpha}_{\mu\nu}$, $\widetilde
{\Gamma}^{\alpha}_{\mu\nu}$ and $\widehat
{\Gamma}^{\alpha}_{\mu\nu}$ to coincide with the Riemannian
connection \ $\overcirc{\Gamma}^{\alpha}_{\mu\nu}$ and the AP-space
in this case becomes trivially a flat Riemannian space .
Consequently, the non-vanishing of any of the three curvature
tensors suffices for the non-vanishing of the torsion tensor.
\vspace{7pt} \pagebreak
\par We now derive some relations that will
prove useful later on. \vspace{-5pt}

\begin{prop} The following relations hold:
\begin{description}
\item[(a)] $\Lambda^{\alpha}_{\mu\nu|\sigma} - \Lambda^{\alpha}_{\mu\nu{\widetilde {|}}\sigma} =
\frak {S}_{\mu,\nu,\sigma} \
\Lambda^{\epsilon}_{\mu\nu}\Lambda^{\alpha}_{\epsilon\sigma}.$
\item[(b)] $\Lambda^{\alpha}_{\mu\nu|\sigma} - \Lambda^{\alpha}_{\mu\nu{\widehat {|}}\sigma} =
\frac{1}{2}(\frak {S}_{\mu,\nu,\sigma} \
\Lambda^{\epsilon}_{\mu\nu}\Lambda^{\alpha}_{\epsilon\sigma}).$
\item[(c)] $\Lambda^{\alpha}_{\mu\nu|\sigma} - \Lambda^{\alpha}_{\mu\nu{o\atop |}\sigma} =
\Lambda^{\epsilon}_{\mu\nu}\gamma^{\alpha}_{\epsilon\sigma} +
\Lambda^{\alpha}_{\nu\epsilon} \gamma^{\epsilon}_{\mu\sigma} +
\Lambda^{\alpha}_{\epsilon\mu}\gamma^{\epsilon}_{\nu\sigma.}$
\end{description}
where the notation $\frak {S}_{\mu,\nu,\sigma}$ denotes a cyclic
permutation of the indices $\mu,\nu,\sigma$ and summation.
\end{prop}

\prof The three relations follow from the definition of the
covariant derivative with respect to the appropriate connection.\ \
$\Box$
\bigskip

Let $M$ be a differentiable manifold equipped with a linear
connection with torsion $T$ and curvature $R$. Then the Bianchi
identities are given locally by \cite{KN}:
\begin{description}
\item []$\frak {S}_{\mu,\nu,\sigma} \ R^{\alpha}_{\mu\nu\sigma} =  \frak {S}_{\mu,\nu,\sigma}  \ (T^{\alpha}_{\mu\nu;\sigma} +
T^{\epsilon}_{\mu\nu} T^{\alpha}_{\epsilon\sigma}), \text { (first
Bianchi identity) }$
\item[] $\frak {S}_{\mu,\nu,\sigma} \ (R^{\alpha}_{\beta \mu\nu;\sigma} +
R^{\alpha}_{\beta \mu\epsilon}T^{\epsilon}_{\nu\sigma}) = 0.\text {
\ \ \ \ \ \ \ \ (second Bianchi identity) }$
\end{description}
where \lq\lq\ ; \rq\rq\, denotes covariant differentiation with
respect to the given connection.
\bigskip

In what follows, we derive some identities using the first and
second Bianchi identities. Some of the derived identities will be
used in simplifying some of the formulae thus obtained.

\begin{thm} The first Bianchi identity for the connections
$\Gamma^{\alpha}_{\mu\nu}$, $\widetilde {\Gamma}^{\alpha}_{\mu\nu}$,
$\widehat {\Gamma}^{\alpha}_{\mu\nu}$ and \
$\overcirc{\Gamma}^{\alpha}_{\mu\nu}$ reads\,{\em:}\vspace{-0.2cm}
\begin{description}
\item[(a)] $\frak {S}_{\mu,\nu,\sigma} \ (\Lambda^{\alpha}_{\mu\nu|\sigma} +
\Lambda^{\epsilon}_{\mu\nu}\Lambda^{\alpha}_{\epsilon\sigma}) = 0.$
\item[(b)] $\frak {S}_{\mu,\nu,\sigma} \ \widetilde {R}^{\alpha}_{\mu\nu\sigma} =
\frak {S}_{\mu,\nu,\sigma} \ (\Lambda^{\alpha}_{\nu\mu\widetilde
{|}\sigma} +
\Lambda^{\epsilon}_{\mu\nu}\Lambda^{\alpha}_{\epsilon\sigma}).$
\item[(c)] $\frak {S}_{\mu,\nu,\sigma} \ \widehat {R}^{\alpha}_{\mu\nu\sigma} = 0.$
\item[(d)] $\frak {S}_{\mu,\nu,\sigma} \ \overcirc{R}^{\alpha}_{\mu\nu\sigma} = 0.$\vspace{-0.2cm}
\end{description}
 {\par \text  The second Bianchi identity for the connections
$\widetilde{\Gamma}^{\alpha}_{\mu\nu}$, $\widehat
{\Gamma}^{\alpha}_{\mu\nu}$ and \
$\overcirc{\Gamma}^{\alpha}_{\mu\nu}$ reads\,{\em:}}\vspace{-0.2cm}
\begin{description}
\item[(e)] $\frak  {S}_{\mu,\nu,\sigma} \ \widetilde {R}^{\alpha}_{\beta\mu\nu\widetilde
{|}\sigma} = \frak {S}_{\mu,\nu,\sigma} \
\Lambda^{\epsilon}_{\sigma\nu}
\Lambda^{\alpha}_{\epsilon\mu|\beta}.$
\item[(f)] $\frak {S}_{\mu,\nu,\sigma} \ \widehat {R}^{\alpha}_{\beta\mu
\nu\widehat {|}\sigma} = 0.$
\item[(g)] $\frak {S}_{\mu,\nu,\sigma} \ \overcirc{R}^{\alpha}_{\beta\mu
\nu{o\atop |}\sigma} = 0.$
\end{description}
\end{thm}

\prof  Identities (a) and (b) follow respectively from the fact that
$R^{\alpha}_{\mu\nu\sigma}$ vanishes identically and that
$\widetilde {\Lambda}^{\alpha}_{\mu\nu} =
\Lambda^{\alpha}_{\nu\mu}$. Identity (e) results from Theorem 3.1
(a) together with the fact that $\widetilde
{\Lambda}^{\alpha}_{\mu\nu} = \Lambda^{\alpha}_{\nu\mu}$. The
remaining identities are trivial because of the symmetry of the
connections $\widehat {\Gamma}^{\alpha}_{\mu\nu}$ and \
$\overcirc{\Gamma}^{\alpha}_{\mu\nu}$.\ \ $\Box$

\begin{prop} The following identities hold:
\begin{description}
\item[(a)] $\frak {S}_{\mu,\nu,\sigma} \ \Lambda^{\alpha}_{\mu\nu{\widetilde {|}}\sigma} = 0.$
\item[(b)] $\frak {S}_{\mu,\nu,\sigma} \
\widetilde{R}^{\alpha}_{\mu\nu\sigma} = \frak {S}_{\mu,\nu,\sigma}
(\Lambda^{\epsilon}_{\mu\nu}\Lambda^{\alpha}_{\epsilon\sigma}).$
\end{description}
\end{prop}

\prof Taking into account Theorem 3.4 (b) together with Theorem 3.1
(a) we get\vspace{-0.2cm}
$$\frak {S}_{\mu,\nu,\sigma} \ (\Lambda^{\alpha}_{\nu\sigma|\mu} +
\Lambda^{\epsilon}_{\mu\nu}\Lambda^{\alpha}_{\epsilon\sigma}) =
\frak {S}_{\mu,\nu,\sigma} \ \Lambda^{\alpha}_{\mu\nu{\widetilde
{|}}\sigma}.\vspace{-0.2cm}$$ By Theorem 3.4 (a), the left hand side
of the the above equation vanishes and the first identity follows.
The second identity is a direct consequence of theorem 3.4 (b),
taking into consideration identity (a) above.\ \ $\Box$
\bigskip

The next result is crucial in simplifying some of the identities
obtained so far and in proving other interesting results.

\begin{thm} The torsion tensor satisfies the identity
$$\frak {S}_{\mu,\nu,\sigma} \ \Lambda^{\epsilon}_{\mu\nu}\Lambda^{\alpha}_{\epsilon\sigma} = 0.$$
\end{thm}

\prof  Applying the first Bianchi identity to $\widehat
R^{\alpha}_{\mu\nu\sigma}$ as expressed in Theorem 3.1 (b), we
get\vspace{-0.2cm}
$$\frac{1}{2}\frak {S}_{\mu,\nu,\sigma} \ (\Lambda^{\alpha}_{\mu\nu|\sigma} - \Lambda^{\alpha}_{\mu\sigma|\nu})
+ \frac{1}{4}\frak {S}_{\mu,\nu,\sigma} \
(\Lambda^{\epsilon}_{\mu\nu}\Lambda^{\alpha}_{\sigma\epsilon} -
\Lambda^{\epsilon}_ {\mu\sigma}\Lambda^{\alpha}_{\nu\epsilon}) +
\frac{1}{2}\frak {S}_{\mu,\nu,\sigma} \
\Lambda^{\epsilon}_{\nu\sigma} \Lambda^{\alpha}_{\mu\epsilon} = 0. \
\ \ (*)\vspace{-0.2cm}$$ Considering each of the above three terms
separately, and taking into account Theorem 3.4 (a), we obtain
respectively
$$\frac{1}{2}\frak {S}_{\mu,\nu,\sigma} \ (\Lambda^{\alpha}_{\mu\nu|\sigma} -
\Lambda^{\alpha}_{\mu\sigma|\nu})  = \frac{1}{2} \frak
{S}_{\mu,\nu,\sigma} \ (\Lambda^{\alpha}_{\mu\nu|\sigma} +
\Lambda^{\alpha}_{\sigma\mu|\nu}) = \frak {S}_{\mu,\nu,\sigma} \
\Lambda^{\alpha}_{\mu\nu|\sigma} = \frak {S}_{\mu,\nu,\sigma} \
\Lambda^{\epsilon}_{\mu\nu}\Lambda^{\alpha}_ {\sigma\epsilon},$$
$$\frac{1}{4}\frak {S}_{\mu,\nu,\sigma} \ (\Lambda^{\epsilon}_{\mu\nu}\Lambda^{\alpha}_
{\sigma\epsilon} -
\Lambda^{\epsilon}_{\mu\sigma}\Lambda^{\alpha}_{\nu\epsilon})
 =  \frac{1}{4}\frak {S}_{\mu,\nu,\sigma} \ (\Lambda^{\epsilon}_
{\mu\nu}\Lambda^{\alpha}_{\sigma\epsilon} + \Lambda^{\epsilon}
_{\sigma\mu}\Lambda^{\alpha}_{\nu\epsilon}) = \frac{1}{2}\frak
{S}_{\mu,\nu,\sigma} \ \Lambda^{\epsilon}_{\mu\nu}\Lambda^{\alpha}_
{\sigma\epsilon}$$ and
$$\frac{1}{2}\frak {S}_{\mu,\nu,\sigma} \ \Lambda^{\epsilon}_{\nu\sigma}\Lambda^{\alpha}_{\mu\epsilon} =
\frac{1}{2}\frak {S}_{\mu,\nu,\sigma} \
\Lambda^{\epsilon}_{\mu\nu}\Lambda^{\alpha}_{\sigma\epsilon}.$$ The
required identity results by substituting the above three equations
into $(*)$.\ \ $\Box$

\begin{cor} The following identities hold:
\begin{description}
\item[(a)] $\Lambda^{\alpha}_{\mu\nu|\sigma} = \Lambda^{\alpha}_
{\mu\nu{\widetilde {|}}\sigma} = \Lambda^{\alpha}_{\mu\nu{\widehat
{|}}\sigma}.$
\item[(b)] $\frak {S}_{\mu,\nu,\sigma} \ \Lambda^{\alpha}_{\mu\nu|\sigma} = 0.$
\item[(c)] $\frak {S}_{\mu,\nu,\sigma} \ \widetilde {R}^{\alpha}_{\mu\nu\sigma} = 0.$
\item[(d)] $\frak{S}_{\mu,\nu,\sigma,} \ \Lambda^{\alpha}_{\nu\mu\widetilde {|}\beta\sigma} =
\frak {S}_{\mu,\nu,\sigma} \
\Lambda^{\epsilon}_{\sigma\nu}\Lambda^{\alpha}_
{\epsilon\mu\widetilde{|}\beta}$
 \ $(\Lambda^{\alpha}_{\nu\mu\widetilde{|}\beta\sigma} : =
\Lambda^{\alpha}_{\nu\mu\widetilde{|}\beta\widetilde{|}\sigma}).$
\item[(e)] $\frak {S}_{\mu,\nu,\sigma} \ (\gamma^{\alpha}_{\mu\nu|\sigma} -
\gamma^{\alpha}_{\nu\mu|\sigma}) = 0.$
\end{description}
\end{cor}

\prof Taking into account Theorem 3.6, relation (a) follows from
Proposition 3.3 (a) and (b), whereas identities (b) and (c) follow
from Theorem 3.4 (a) and Proposition 3.5 (b) respectively. Identity
(d) follows from Theorem 3.4 (e) taking into account Theorem 3.1 (a)
together with relation (a) above. Finally, the last identity follows
from identity (b) above together with the relation
$\Lambda^{\alpha}_{\mu\nu} = \gamma^{\alpha}_{\mu\nu} -
\gamma^{\alpha}_{\nu\mu}$.\ \ $\Box$
\bigskip

In view of Theorem 3.6, the curvature tensor $\widehat
R^{\alpha}_{\mu\nu\sigma}$ can be further simplified as revealed in

\begin{prop} The curvature tensor $\widehat {R}^{\alpha}_{\mu\nu\sigma}$ can be
expressed in the form:
$$\widehat {R}^{\alpha}_{\mu\nu\sigma} = \frac{1}{2} \Lambda^{\alpha}_{\sigma\nu|\mu} +
\frac{1}{4}
\Lambda^{\epsilon}_{\sigma\nu}\Lambda^{\alpha}_{\epsilon\mu}.$$
\end{prop}

\prof  The curvature tensor $\widehat R^{\alpha}_{\mu\nu\sigma}$ has
the form (Theorem 3.1 (b)):
$$\widehat {R}^{\alpha}_{\mu\nu\sigma} =
\frac{1}{2}(\Lambda^{\alpha}_{\mu\nu|\sigma} -
\Lambda^{\alpha}_{\mu\sigma|\nu}) +
\frac{1}{4}(\Lambda^{\epsilon}_{\mu\nu}\Lambda^{\alpha}_{\sigma\epsilon}
-
 \Lambda^{\epsilon}_{\mu\sigma}\Lambda^{\alpha}_{\nu\epsilon}) +
\frac{1}{2}(\Lambda^{\epsilon}_{\sigma\nu}\Lambda^{\alpha}_{\epsilon\mu}).$$
Taking into account identity (b) in Corollary 3.7 and Theorem 3.6,
we get
\begin{eqnarray*}
\widehat {R}^{\alpha}_{\mu\nu\sigma} &=& \frac{1}{2}\{(\frak
{S}_{\mu,\nu,\sigma} \ \Lambda^{\alpha}_{\mu\nu|\sigma}) -
\Lambda^{\alpha}_{\nu\sigma|\mu}\}
 + \frac{1}{2}\Lambda^{\epsilon}_{\nu\sigma}\Lambda^{\alpha}_{\mu\epsilon}\\
&& + \ \frac{1}{4}\{(\frak {S}_{\mu,\nu,\epsilon} \
\Lambda^{\epsilon}_{\mu\nu}\Lambda^{\alpha}_{\sigma\epsilon}) -
\Lambda^{\epsilon}_{\nu\sigma} \ \Lambda^{\alpha}_{\mu\epsilon}\}\\
&=& \frac{1}{2}\Lambda^{\alpha}_{\sigma\nu|\mu} +
\frac{1}{4}\Lambda^{\epsilon}_{\sigma\nu}
\Lambda^{\alpha}_{\epsilon\mu}.\ \ \Box
\end{eqnarray*}

\bigskip
The formula obtained in Theorem 3.1 for the curvature tensor
$\widetilde R^{\alpha}_{\mu\nu\sigma}$ is strikingly compact and
elegant. The formula obtained for $\widehat
{R}^{\alpha}_{\mu\nu\sigma}$ is however less elegant but still
relatively compact.  These two formulae, together with the first
Bianchi identity, enabled us to derive, in a simple way, the crucial
identities $\frak{S}_{\mu,\nu,\sigma} \
\Lambda^{\alpha}_{\mu\nu|\sigma} = 0$ and $\frak{S}_{\mu,\nu,\sigma}
\ \Lambda^{\epsilon}_{\mu\nu}\Lambda^{\alpha}_{\epsilon\sigma} = 0$
which, in turn, simplified the formula obtained for $\widehat
R^{\alpha}_{\mu\nu\sigma}$ (which is now more elegant). The last two
identities will play an essential role in the rest of the paper. It
should be noted however that a direct proof of these identities is
far from clear.
\begin{rem} All results and identities concerning the connections
$\Gamma^{\alpha}_{\mu\nu}$, $\widetilde {\Gamma}^{\alpha}_{\mu\nu}$
and $\widehat {\Gamma}^{\alpha}_{\mu\nu}$ {\em({\it resp. the
connection $\overcirc{\Gamma}^{\alpha}_{\mu\nu}$})} remain valid in
the more general context in which $\Gamma^{\alpha}_{\mu\nu}$ is a
non-symmetric linear connection on a manifold $M$ {\em({\it resp. a
Riemannian manifold $(M, \ g)$})} with vanishing curvature.
\end{rem}


\Section{The Wanas Tensor (W-tensor)}

Let $(M,\lambda)$ be an AP-space of dimension $n$, where $\lambda$
denotes any one of the $n$ linearly independent vector fields
defining the AP-structure on the manifold $M$. Let
${\widetilde{R}}^{\alpha}_{\mu\nu\sigma}$ and $\widetilde
{\Lambda}^{\alpha}_{\mu\nu}$ be the curvature and the torsion
tensors of the dual connection $\widetilde
{\Gamma}^{\alpha}_{\mu\nu}$.

\begin{defn} The tensor field $\,W^{\alpha}_{\mu\nu\sigma}$ of type (1,3) on $M$
defined by the formula
$$\lambda_{\epsilon}W^{\epsilon}_{\mu\nu\sigma} :=
\lambda_{\mu\widetilde{|}\nu\sigma} - \
\lambda_{\mu\widetilde{|}\sigma\nu}$$ will be called the Wanas
tensor, or simply the W-tensor, of the AP-space $(M,\lambda)$.
\end{defn}

The W-tensor  $\,W^{\alpha}_{\mu\nu\sigma}\,$  has been first
defined by M. I Wanas \cite{a} in 1975 and has been used by F.
Mikhail and M. Wanas \cite{MW} to construct a pure geometric theory
unifying gravity and electromagnetism. A double contraction of the
tensor $\,g^{\mu\gamma}\, W^{\alpha}_{\mu\nu\sigma}\,$ gives a
scalar that has been used to write the Lagrangian density for fields
and matter. The symmetric part of the field equations obtained
contains a second order tensor representing the material
distribution. This tensor is a pure geometric object and not a
phenomenological one. The skew part of the field equations gives
rise to Maxwell-like equations. In this theory the metric tensor
plays the role of the gravitational potential while the the vector
$c_{\mu}$ (the contracted torsion) plays the role of the
electromagnetic potential. The linearized form \cite{MW1} of the
theory supports these identifications.

The next result gives quite a simple expression for such a tensor.

\begin{thm} Let $(M,\lambda)$ be an AP-space. Then the W-tensor can
be expressed in the form
$$W^{\alpha}_{\mu\nu\sigma} = \Lambda^{\alpha}_{\sigma\nu|\mu} -
\Lambda^{\epsilon}_{\sigma\nu} \Lambda^{\alpha}_{\epsilon\mu},$$
where $\Lambda^{\alpha}_{\mu\nu}$ is the torsion tensor of the
canonical connection $\Gamma^{\alpha}_{\mu\nu}$.
\end{thm}

\prof  Consider the commutation formula with respect to the
connection $\widetilde {\Gamma}^{\alpha}_{\mu\nu}$:
$$\lambda_{\mu\widetilde{|}\nu\sigma} - \ \lambda_
{\mu\widetilde{|}\sigma\nu} = \lambda_{\epsilon} \widetilde
{R}^{\epsilon}_{\mu\nu\sigma} + \lambda_{\mu\widetilde{|}\epsilon}
\widetilde {\Lambda}^{\epsilon}_{\nu\sigma}.$$ Multiplying both
sides by $\lambda^{\alpha}$, using the definition of the W-tensor
together with the definition of $\Gamma^{\alpha}_{\mu\nu}$ and
taking into account Theorem 3.1 (a), we get
\begin{eqnarray*} W^{\alpha}_{\mu\nu\sigma} &=&
\Lambda^{\alpha}_{\sigma\nu|\mu} + \undersym{\lambda}{i}^{\alpha}
(\undersym{\lambda}{i}_{\mu, \epsilon} -
\undersym{\lambda}{i}_{\beta}\Gamma^{\beta}_{\epsilon\mu})
\Lambda^{\epsilon}_{\sigma\nu}\\
&=& \Lambda^{\alpha}_{\sigma\nu|\mu} +
(\Gamma^{\alpha}_{\mu\epsilon} -
\Gamma^{\alpha}_{\epsilon\mu})\Lambda^{\epsilon}_{\sigma\nu}\\
&=& \Lambda^{\alpha}_{\sigma\nu|\mu} -
\Lambda^{\epsilon}_{\sigma\nu}\,\Lambda^{\alpha}_{\epsilon\mu}.\ \
\Box
\end{eqnarray*}
\begin{rem} The W-tensor can be also defined contravariantly as follows:
$$\lambda^{\mu}\,W^{\alpha}_{\mu\nu\sigma} =
\lambda^{\alpha}\,_{\widetilde{|}\nu\sigma} - \lambda^{\alpha}\,
_{\widetilde{|}\sigma\nu} = \lambda^{\epsilon} \widetilde
{R}^{\alpha}_{\epsilon\sigma\nu} + \lambda^{\alpha}\,
_{\widetilde{|}\epsilon} \,
\widetilde{\Lambda}^{\epsilon}_{\nu\sigma}.$$
\end{rem}
This definition gives the same formula for the W-tensor as in
Theorem 4.2.
\begin{prop} The Wanas tensor has the following properties:
\begin{description}
\item[a)] $W^{\alpha}_{\mu\nu\sigma}$ is skew symmetric in the last pair of indices.
\item[(b)] $W^{\alpha}_{\mu\nu\sigma|\beta} - W^{\alpha}_{\mu\nu\sigma\widetilde{|}
\beta} = \Lambda^{\alpha}_{\sigma\nu|\mu\beta} -
\Lambda^{\alpha}_{\sigma\nu\widetilde{|}\mu\beta}.$
\end{description}
\end{prop}

\prof The first property is trivial. The second property holds as a
result of Theorem 4.2 together with corollary 3.7 (a). \ \ $\Box$

\begin{thm} The W-tensor satisfies the following identity:
$$\frak{S}_{\mu,\nu,\sigma} \ W^{\alpha}_{\mu\nu\sigma} = 0.$$
\end{thm}

\prof Follows directly from Theorem 4.2, taking into account Theorem
3.6 together with Corollary 3.7 (b).\ \ $\Box$
\bigskip

The identity satisfied by the W-tensor in Theorem 4.5 is the same as
the first Bianchi identity of the Riemannian curvature tensor. The
identity corresponding to the second Bianchi identity is given by:
\begin{thm} The W-tensor satisfies the following identity:
$$\frak{S}_{\nu,\sigma,\beta} \ W^{\alpha}_{\mu\nu\sigma\widetilde{|}\beta} =
\frak{S}_{\nu,\sigma,\beta} \ \{\Lambda^{\epsilon}_
{\nu\sigma}(\Lambda^{\alpha}_{\epsilon\beta|\mu}  +
\Lambda^{\alpha}_{\epsilon\mu|\beta})\}.$$
\end{thm}

\prof Taking into account the second Bianchi identity, Theorem 4.2,
Theorem 3.1 (a) and Corollary 3.8 (a) and (b), we get
\begin{eqnarray*} \frak{S}_{\nu,\sigma,\beta} \ W^{\alpha}_{\mu\nu\sigma\widetilde{|}\beta}
&=& \frak{S}_{\nu,\sigma,\beta} \ \{\widetilde {R}^{\alpha}_
{\mu\nu\sigma\widetilde{|}\beta} -
(\Lambda^{\epsilon}_{\sigma\nu}\Lambda^{\alpha}_
{\epsilon\mu})_{\widetilde{|}\beta}\}\\
&=& \frak{S}_{\nu,\sigma,\beta} \ (\widetilde
{R}^{\alpha}_{\mu\nu\epsilon}\widetilde
{\Lambda}^{\epsilon}_{\beta\sigma} -
\Lambda^{\epsilon}_{\sigma\nu}\Lambda^{\alpha}_{\epsilon\mu|\beta})\\
&=& \frak{S}_{\nu,\sigma,\beta} \
(\Lambda^{\alpha}_{\epsilon\nu|\mu} \Lambda^{\epsilon}_{\sigma\beta}
- \Lambda^{\alpha}_{\epsilon\mu|\beta}
\Lambda^{\epsilon}_{\sigma\nu})\\
&=& \frak{S}_{\nu,\sigma,\beta} \ \{\Lambda^{\epsilon}_{\nu\sigma}
(\Lambda^{\alpha}_{\epsilon\beta|\mu} +
\Lambda^{\alpha}_{\epsilon\mu|\beta})\}. \ \ \Box
\end{eqnarray*}
We conclude this section by some comments\,:
\begin{itemize}
\item
The W-tensor is defined in terms of the dual connection
$\widetilde{\Gamma}^{\alpha}_{\mu\nu}$. The same definition using
the three other connections gives nothing new. In fact, the
commutation formula for the connection $\Gamma^{\alpha}_{\nu\mu}$ is
trivial (since $\lambda^{\alpha}\,_{|\mu} = 0$), whereas the
commutation formulae for $\widehat {\Gamma}^{\alpha}_{\mu\nu}$ and \
$\overcirc{\Gamma}^{\alpha}_{\mu\nu}$ give rise to $\widehat
{R}^{\alpha}_{\nu\mu\sigma}$ and \
$\overcirc{R}^{\alpha}_{\mu\nu\sigma}$ respectively (since the
torsion tensor vanishes in the latter two cases).
\item
The vanishing of the torsion tensor implies the vanishing of the
W-tensor which is equivalent to the commutativity of successive
covariant differentiation (of the vector fields $\lambda$); a
striking property which does not exist in Riemannian geometry (or
even in other geometries, in general).
\item
The W-tensor has some properties common with the Riemannian
curvature tensor (for example, Proposition 4.4 (a) and Theorem 4.5).
Nevertheless, there are significant deviations from the Riemannian
curvature tensor (for example, Theorem 4.6 ).
\item
As stated above, Mikhail and Wanas \cite{MW}, in their theory, have
attributed the gravitational potential to the metric tensor and the
electromagnetic potential to the contracted torsion. The use of the
$W$-tensor has aided to construct such a theory \textbf{via one
single geometric entity}, which Einstein was seeking for \cite{En1}.
Different applications (for example, \cite{c}, \cite{d}) of
Mikhail-Wanas theory support these statements. In conclusion, the
expression of the W-tensor comprises, in addition to the curvature
tensor $\widetilde {R}^{\alpha}_{\mu\nu\sigma}$, a term containing a
torsion contribution. Thus, the W-tensor expresses {\it
geometrically} the interaction between curvature and torsion. On the
other hand, as gravity is described in terms of curvature and
electromagnetism is described in terms of torsion, the W-tensor
expresses {\it physically} the interaction between gravity and
electromagnetism.
\end{itemize}

\Section{The semi-symmetric case} \vspace{-0.2cm} In this section,
we investigate an AP-space whose canonical connection has a special
simple form.
\begin{defn} Let $M$ be a differentiable manifold. A semi-symmetric
connection on $M$ is a linear connection on $M$ whose torsion tensor
$T$ is given by {\em \cite{KY}}\vspace{-0.2cm}
$$T^{\alpha}_{\mu\nu} = \delta^{\alpha}_{\mu} w_{\nu} - \delta^{\alpha}_{\nu} w_{\mu},\vspace{-0.2cm}$$
for some scalar 1-form $w_{\mu}$.
\end{defn}

Semi-symmetric connections have been studied by many authors (cf.
for example \cite{AS}, \cite{T}, \cite{Y}, \cite{LY}). In what
follows, we consider an $n$-dimensional AP-space $(M,\lambda)$ with
the additional assumption that the canonical connection
$\Gamma^{\alpha}_{\mu\nu}$ is semi-symmetric. Then, by Definition
5.1, we have
$$\Lambda^{\alpha}_{\mu\nu} =
\delta^{\alpha}_{\mu} w_{\nu} - \delta^{\alpha}_{\nu} w_{\mu}.$$
Moreover, it can be shown that \cite{TI}
$$\Gamma^{\alpha}_{\mu\nu} = \overcirc{\Gamma}^{\alpha}_{\mu\nu} + \delta^{\alpha}_{\mu}w_{\nu} -
g_{\mu\nu} w^{\alpha},$$ where \
$\overcirc{\Gamma}^{\alpha}_{\mu\nu}$ is the Riemannian connection
defined in section 2 and $w^{\alpha}= g^{\alpha\mu}w_{\mu}.$ Hence,
$$\gamma^{\alpha}_{\mu\nu} = \delta^{\alpha}_{\mu} w_{\nu} - g_{\mu\nu} w^{\alpha}.$$

\begin{prop} The curvature tensors $\widetilde {R}^{\alpha}_{\mu\nu\sigma}$,
$\widehat {R}^{\alpha}_{\mu\nu\sigma}$ and \ $\overcirc
{R}^{\alpha}_{\mu\nu\sigma}$ of the connections $\widetilde
{\Gamma}^{\alpha}_{\mu\nu}$, $\widehat {\Gamma}^{\alpha}_{\mu\nu}$
and \ $\overcirc{\Gamma}^{\alpha}_{\mu\nu}$ are expressed in terms
of $w_{\mu}$ in the form{\em:}
\begin{description}
\item[(a)] $\widetilde{R}^{\alpha}_{\mu\nu\sigma} = \delta^{\alpha}_{\sigma}
w_{\nu|\mu} -\delta^{\alpha}_{\nu} w_{\sigma|\mu}.$
\item[(b)] $\widehat {R}^{\alpha}_{\mu\nu\sigma} =
\frac{1}{2}(\delta^{\alpha}_{\sigma}w_{\nu|\mu} -
\delta^{\alpha}_{\nu} w_{\sigma|\mu}) + \frac{1}{4}
w_{\mu}(\delta^{\alpha}_{\sigma}w_{\nu} - \delta^{\alpha}_{\nu}
w_{\sigma}).$
\item[(c)] $\overcirc{R}^{\alpha}_{\mu\nu\sigma} = \delta^{\alpha}_{\mu}
(w_{\nu|\sigma} - w_{\sigma|\nu})
+ (g_{\mu\sigma} w^{\alpha} \ _{|\nu} - g_{\mu\nu} w^{\alpha} \
_{|\sigma})  + 2w^{\alpha}(g_{\mu\sigma}w_{\nu} -
g_{\mu\nu}w_{\sigma}).$
\end{description}
Consequently, if $\widetilde {R}_{\alpha\mu\nu\sigma} :=
g_{\alpha\epsilon}\widetilde {R}^{\epsilon}_{\mu\nu\sigma}$ with
similar expressions for $\widehat R_{\alpha\mu\nu\sigma}$ and \
$\overcirc{R}_{\alpha\mu\nu\sigma},$ we get{\em:}
\begin{description}
\item[(\'{a})]
$ \widetilde{R}_{\alpha\mu\nu\sigma} = g_{\alpha\sigma} w_{\nu|\mu}
- g_{\alpha\nu} w_{\sigma|\mu}.$

\item[(\'{b})] $\widehat {R}_{\alpha\mu\nu\sigma} = \frac{1}{2}(g_{\alpha\sigma}
w_{\nu|\mu} - g_{\alpha\nu} w_{\sigma|\mu}) + \frac{1}{4}
w_{\mu}(g_{\alpha\sigma}w_{\nu} - g_{\alpha\nu} w_{\sigma}).$

\item[(\'{c})] $\overcirc{R}_{\alpha\mu\nu\sigma} = g_{\alpha\mu}(w_{\nu|\sigma} -
w_{\sigma|\nu}) + (g_{\mu\sigma} w_{\alpha|\nu} - g_{\mu\nu}
w_{\alpha|\sigma})  + 2w_{\alpha}(g_{\mu\sigma}w_{\nu} -
g_{\mu\nu}w_{\sigma}).$
\end{description}
\end{prop}

\prof The first two relations hold by substituting
$\Lambda^{\alpha}_{\mu\nu}$ in the formulae expressing
$\widetilde{R}^{\alpha}_{\mu\nu\sigma}$ (Theorem 3.1 (a)) and
$\widehat {R}^{\alpha}_{\mu\nu\sigma}$ (Proposition 3.8)
respectively. The third relation holds by substituting
$\Lambda^{\alpha}_{\mu\nu}$ and $\gamma^{\alpha}_{\mu\nu}$ in the
formula expressing \ $\overcirc{R}^{\alpha}_{\mu\nu\sigma}$ (Theorem
3.1 (c)). The remaining relations are straightforward.\ \ $\Box$

\pagebreak
\begin{prop} Let $\widetilde{R}_{\mu\nu} := \widetilde {R}^{\alpha}_{\mu\nu\alpha}$
and $\widetilde{R} := g^{\mu\nu}\widetilde{R}_{\mu\nu}$, with \
similar \ expressions \ for \ $\widehat R^{\alpha}_{\mu\nu\sigma}$
and \ $\overcirc{R}^{\alpha}_{\mu\nu\alpha}$. Then
\begin{description}
\item[(a)] $\widetilde{R}_{\mu\nu} = (n - 1)w_{\nu|\mu},
\ \widetilde {R} = (n - 1)w^{\mu} \ _{|\mu}.$
\item[(b)] $\widehat {R}_{\mu\nu} = \frac{1}{4}(n - 1)(2w_{\nu|\mu} + w_{\mu}w_{\nu}),\
\widehat {R} = \frac{1}{4}(n - 1)(2w^{\mu} \ _{|\mu} +
w^{\mu}w_{\mu}).$
\item[(c)] $\overcirc{R}_{\mu\nu} = w_{\nu|\mu} - g_{\mu\nu} w^{\sigma}\ _{|\sigma} +
2(w_{\nu}w_{\mu} - g_{\mu\nu}w^{\sigma}w_{\sigma}), \ \,
\overcirc{R} = (1 - n)({w^{\mu}\ _{|\mu}} + 2{w^{\mu}w_{\mu}}).$
\end{description}
\end{prop}

\prof  Follows directly from the relations obtained in Proposition
5.2 by applying the suitable contractions.\ \ $\Box$

\begin{thm} The second order covariant tensor $w_{\nu|\mu}$ is
symmetric:
$$w_{\mu|\nu} = w_{\nu|\mu}.$$
\end{thm}

\prof  Follows directly from Proposition 5.3 (c) noting that \
$\overcirc{R}_{\mu\nu}$ is symmetric, being the Ricci tensor of the
Riemannian connection. This can be also deduced from Theorem 5.2 (c)
noting that \ $\overcirc{R}^{\mu}_{\mu\nu\sigma} = 0.$\ \ $\Box$

\bigskip
A direct consequence of the above theorem is the following
\begin{thm} For an AP-space $(M,\lambda)$ with semi-symmetric connection
$\Gamma^{\alpha}_{\mu\nu}$, the following properties hold:
\begin{description}
\item [(a)]$\widetilde{R}_{\mu\nu}$ and $\widehat R_{\mu\nu}$ are symmetric.
\item [(b)]$\widetilde {R}^{\mu}_{\mu\nu\sigma} = 0.$
\item [(c)]$\widehat {R}^{\mu}_{\mu\nu\sigma} = 0.$
\end{description}
\end{thm}

\begin{thm} For an AP-space $(M,\lambda)$ with semi-symmetric connection
$\Gamma^{\alpha}_{\mu\nu}$, the W-tensor has the form:
\begin{description}
\item[(a)] $W^{\alpha}_{\mu\nu\sigma} = \delta^{\alpha}_{\sigma}(w_{\nu|\mu} - w_{\mu}w_{\nu}) -
\delta^{\alpha}_{\nu}(w_{\sigma|\mu} - w_{\sigma}w_{\mu}).$
\item[(b)]$W_{\alpha\mu\nu\sigma} = g_{\alpha\sigma}(w_{\nu|\mu} - w_{\mu}w_{\nu}) -
g_{\alpha\nu}(w_{\sigma|\mu} - w_{\sigma}w_{\mu})$ \ \ \
{\em{(}}$W_{\alpha\mu\nu\sigma} =
g_{\alpha\epsilon}W^{\epsilon}_{\mu\nu\sigma}${\em{)}}.
\end{description}
\end{thm}
\noindent{\bf{Proof.}} Follows directly from Theorem 4.2 by
substituting the expression of $\Lambda^{\alpha}_{\mu\nu}$ in terms
$w_{\mu}$.\ \ $\Box$
\begin{cor} Let $W_{\mu\nu} := W^{\epsilon}_{\mu\nu\epsilon}$ and
$W := g^{\mu\nu}W_{\mu\nu}$. Then
\begin{description}
\item[(a)] $W_{\mu\nu} = (n - 1)(w_{\nu|\mu} - w_{\mu} w_{\nu}).$
\item[(b)]  $W = (n - 1)(w^{\mu} \ _{|\mu} - w ^{\mu}w_{\mu}).$
\item[(c)] $W^{\mu}_{\mu\nu\sigma} = 0.$

Consequently, $W_{\mu\nu}$ is symmetric.
\end{description}
\end{cor}

Comparing Corollary 5.7 (a) and Theorem 5.6 (a), we obtain
\begin{thm} Let $dim \ M\geq 2$. A sufficient condition for the vanishing of the W-tensor
$\,W^{\alpha}_{\mu\nu\sigma}\,$ is that $W_{\mu\nu} = 0$.
\end{thm}
Finally, in view of Theorem 1 of K. Yano \cite{KY}, the following
result follows:
\begin{thm} If the canonical connection $\Gamma^{\alpha}_{\mu\nu}$
of an AP-space is semi-symmetric, then the associated Riemannian
metric $g_{\mu\nu}$ is conformally flat.
\end{thm}

\bigskip
{\bf Special case.} In the following we assume that the canonical
connection $\Gamma^{\alpha}_{\mu\nu}$ is a semi-symmetric connection
whose defining 1-form $w_{\mu}$ satisfies the condition
$$w_{\mu}w^{\nu} = \delta^{\nu}_{\mu}.$$
It is easy to see that the above condition implies that $w_{\mu|\nu}
= 0$ and that $w_{\mu}w_{\nu} = g_{\mu\nu}$. Under the given
assumption, the different curvature tensors and the Wanas tensor,
according to Proposition 5.2 and Theorems 5.6, take the form:
$$\widetilde {R}^{\alpha}_{\mu\nu\sigma} = 0,$$
$$\widehat {R}^{\alpha}_{\mu\nu\sigma} = \frac{1}{4}
(\delta^{\alpha}_{\sigma} \, g_{\mu\nu} - \delta^{\alpha}_{\nu}\,
g_{\mu\sigma}),$$
$$\overcirc{R}^{\alpha}_{\mu\nu\sigma} =
2(\delta^{\alpha}_{\nu} \, g_{\mu\sigma} - \delta^{\alpha}_{\sigma}
\, g_{\mu\nu}),$$

$$W^{\alpha}_{\mu\nu\sigma} = \delta^{\alpha}_{\nu} \, g_{\mu\sigma} -
\delta^{\alpha}_{\sigma} \, g_{\mu\nu}.$$ Consequently,
$$W^{\alpha}_{\mu\nu\sigma} = - 4\widehat {R}^{\alpha}_{\mu\nu\sigma} =
\frac{1}{2} \ \overcirc{R}^{\alpha}_{\mu\nu\sigma}$$ and
$$\overcirc{R}_{\alpha\mu\nu\sigma} = 2(g_{\alpha\nu}g_{\mu\sigma} -
g_{\alpha\sigma}g_{\mu\nu}).$$

In this case, the W-tensor becomes a curvature-like tensor \cite{KN}
and the above formulae imply the following result:

\begin{thm} Let the canonical connection $\Gamma^{\alpha}_{\mu\nu}$ of an AP-space
$(M,\lambda)$ be a semi-symmetric connection whose defining 1-form
$w_{\mu}$ satisfies the condition
$$w_{\mu}w^{\nu} = \delta^{\nu}_{\mu}.$$
Then, all nonvanishing curvature tensors of $(M,\lambda)$ coincide,
up to a constant, with the W-tensor and the AP-space becomes a
Riemannian space of constant curvature.
\end{thm}

It should be noted that, in this particular case, the AP-character
of the space recedes, whereas the Riemannian aspects of the AP-space
become dominant. Physically speaking, the latter result seems to
suggest that, in this particular case, electromagnetic effects are
absent. This particular case can thus be considered, in some sense,
as a limiting case.

\bibliographystyle{plain}

\begin{thebibliography}{100}

\bibitem{AS} O. C. Andonie et D. Smaranda, {\it Certaines connexions
semi-sym\'{e}trique}, Tensor, N.S., Vol. 31 (1977), 8-12.

\bibitem{BC} F. Brickell and R. S. Clark, {\it Differentiable manifolds},
Van Nostrand Reinhold Co., 1970.

\bibitem{En} A. Einstein, {\it Unified field theory based on Riemannian metrics and
distant parallelism}, Math. Annal., 102 (1930), 685-697.

\bibitem{En1} A. Einstein, {\it The meaning of relativity}, 5th. Ed.,
Princeton Univ. Press, 1955.

\bibitem{HS} K. Hayashi and T. Shirafuji, {\it New general relativity}, Phys. Rev.
D 19 (1979), 3524-3554.

\bibitem {T} T. Imai, {\it Notes on semi-symmetric metric connections},
Tensor, N.S., Vol. 24 (1972), 293-296.

\bibitem {TI} T. Imai, {\it Notes on semi-symmetric metric connections},
Tensor, N.S., Vol. 27 (1973), 56-58.

\bibitem {KN} S. Kobayashi and K. Nomizu, {\it Foundations of differential geometry},
Vol. I. Interscience Publishers, 1963.

\bibitem {MW} F. I. Mikhail and M.I. Wanas, {\it A generalized field theory, I. Field
equations}, Proc. R. Soc. London, A. 356 (1977), 471-481.

\bibitem {MW1} F. I. Mikhail and M.I. Wanas, {\it A generalized field theory, II. Linearized
field equations}, Int. J. Theoret. Phys., 20 (1981), 671-680.

\bibitem{HP} H. P. Robertson, {\it Groups of motion in spaces admitting absolute
parallelism}, Ann. Math, Princeton (2), 33 (1932), 496 - 520.

\bibitem{Tam1} L. Tam\'{a}ssy, {\it Area and metrical connections in
Finsler spaces, Finsler geometries}, A meeting of Minds, Kluwer
Publ., 2000, Proc. of Coll. on Finsler Spaces, Edmonton, July 1998,
263-281.

\bibitem{Tam2} L. Tam\'{a}ssy, {\it Geometry of the point Finsler
spaces}, From the Book [Proceedings of J. Boliai Memorial
Conference, Budapest, July 2006], {\lq\lq Non-Euclidean
geometries\rq\rq}, Springer, 2006, 445-461.

\bibitem{a} M. I. Wanas, {\it A generalized field theory and its applications in
cosmology}, Ph.D Thesis, Cairo University, 1975.

\bibitem{c} M. I. Wanas, {\it A generalized field theory: Charged spherical symmetric solutions},
Int. J. Theoret. Phys., 24 (1985), 639-651.

\bibitem{d} M. I. Wanas, {\it A self consistent world model},
Astrophys. Space Sci., 154 (1989), 165-177.

\bibitem{b} M. I. Wanas, {\it Absolute parallelism geometry: Developments,
applications and problems}, Stud. Cercet, Stiin. Ser. Mat. Univ.
Bacau, No. 10 (2001), 297-309.

\bibitem{KY} K. Yano, {\it On semi-symmetric metric connection}, Rev. Roumaine Math.
Pures Appl., T. 15 (1970), 1579-1586.

\bibitem{Y} N. L. Youssef, {\it Connexions m\'{e}trique semi-sym\'{e}triques
semi-basiques}, Tensor, N.S., Vol. 40 (1983), 242-248.

\bibitem{LY} N. L. Youssef, {\it Vertical semi-symmetric metric connections},
Tensor, N.S., Vol. 49 (1990), 218-229.

















%















\end{thebibliography}


\end{document}